\documentclass[submission]{eptcs}
\providecommand{\event}{TARK 2017} 

\newtheorem{THEOREM}{Theorem}[section]
\newenvironment{theorem}{\begin{THEOREM} \hspace{-.85em} {\bf :} }%
                        {\end{THEOREM}}
\newtheorem{LEMMA}[THEOREM]{Lemma}
\newenvironment{lemma}{\begin{LEMMA} \hspace{-.85em} {\bf :} }%
                      {\end{LEMMA}}
\newtheorem{COROLLARY}[THEOREM]{Corollary}
\newenvironment{corollary}{\begin{COROLLARY} \hspace{-.85em} {\bf :} }%
                          {\end{COROLLARY}}
\newtheorem{PROPOSITION}[THEOREM]{Proposition}
\newenvironment{proposition}{\begin{PROPOSITION} \hspace{-.85em} {\bf :} }%
                            {\end{PROPOSITION}}
\newtheorem{DEFINITION}[THEOREM]{Definition}
\newenvironment{definition}{\begin{DEFINITION} \hspace{-.85em} {\bf :} \rm}%
                            {\end{DEFINITION}}
\newtheorem{CLAIM}[THEOREM]{Claim}
\newenvironment{claim}{\begin{CLAIM} \hspace{-.85em} {\bf :} \rm}%
                            {\end{CLAIM}}
\newtheorem{EXAMPLE}[THEOREM]{Example}
\newenvironment{example}{\begin{EXAMPLE} \hspace{-.85em} {\bf :} \rm}%
                            {\end{EXAMPLE}}
\newtheorem{REMARK}[THEOREM]{Remark}
\newenvironment{remark}{\begin{REMARK} \hspace{-.85em} {\bf :} \rm}%
                            {\end{REMARK}}

\newcommand{\thm}{\begin{theorem}}
\newcommand{\lem}{\begin{lemma}}
\newcommand{\pro}{\begin{proposition}}
\newcommand{\dfn}{\begin{definition}}
\newcommand{\rem}{\begin{remark}}
\newcommand{\xam}{\begin{example}}
\newcommand{\cor}{\begin{corollary}}
\newcommand{\prf}{\noindent{\bf Proof:} }
\newcommand{\ethm}{\end{theorem}}
\newcommand{\elem}{\end{lemma}}
\newcommand{\epro}{\end{proposition}}
\newcommand{\edfn}{\bbox\end{definition}}
\newcommand{\erem}{\bbox\end{remark}}
\newcommand{\exam}{\bbox\end{example}}
\newcommand{\ecor}{\end{corollary}}
\newcommand{\eprf}{\bbox\vspace{0.1in}}
\newcommand{\beqn}{\begin{equation}}
\newcommand{\eeqn}{\end{equation}}

\newcommand{\bbox}{\vrule height7pt width4pt depth1pt}

\newcommand{\clm}{\begin{claim}}
\newcommand{\eclm}{\end{claim}}

\newcommand{\sat}{\models}

\newcommand{\rimp}{\Rightarrow}

\newcommand{\union}{\cup}
\newcommand{\inter}{\cap}

\newcommand{\IN}{\mbox{$I\!\!N$}}

\renewcommand{\phi}{\varphi}
\newcommand{\Circ}{\mbox{{\small $\bigcirc$}}}

\newcommand{\A}{{\cal A}}

\newcommand{\C}{{\cal C}}

\newcommand{\I}{{\cal I}}

\newcommand{\K}{{\cal K}}

\renewcommand{\P}{{\cal P}}

\newcommand{\R}{{\cal R}}

\newcommand{\U}{{\cal U}}

\newcommand{\Y}{{\cal Y}}

\newcommand{\ol}{\setlength{\itemsep}{0pt}\begin{enumerate}}
\newcommand{\eol}{\end{enumerate}\setlength{\itemsep}{-\parsep}}
\newcommand{\ul}{\setlength{\itemsep}{0pt}\begin{itemize}}
\newcommand{\dl}{\setlength{\itemsep}{0pt}\begin{description}}
\newcommand{\edl}{\end{description}\setlength{\itemsep}{-\parsep}}
\newcommand{\eul}{\end{itemize}\setlength{\itemsep}{-\parsep}}

\newcommand{\cS}{{\cal S}}

\newcommand{\BS}{B^{\scriptscriptstyle \cS}}

\newcommand{\ES}{E_\cS}

\newcommand{\CS}{C_\cS}

\newcommand{\commentout}[1]{}

\newcommand{\bi}{\begin{itemize}}
\newcommand{\ei}{\end{itemize}}
\newcommand{\be}{\begin{enumerate}}
\newcommand{\ee}{\end{enumerate}}

\usepackage{amsfonts}
\usepackage{mathrsfs}
\usepackage{amssymb}
\usepackage{latexsym}
\usepackage{graphicx}
\usepackage{enumerate}
\usepackage{amsmath}
\usepackage{mathtools}
\usepackage{mathrsfs}
\newcommand{\eps}{\varepsilon}

        \newcommand{\intension}[2]{[\![ #2 ]\!]_{ #1}}
        \renewcommand{\H}{{\mathcal{H}}}
\newcommand{\AG}{{\mathcal{AG}}}
        
                  \newcommand{\Tprefix}{{\mathit{T\mbox{-}prefix}}}
\newcommand{\fullv}[1]{}
\newcommand{\shortv}[1]{#1}
\fullv{ \usepackage{chicagor}}

\renewcommand{\R}{{\mathcal R}}
\renewcommand{\I}{{\mathcal I}}
\renewcommand{\cS}{{\mathcal S}}
\renewcommand{\A}{{\mathcal A}}
\renewcommand{\K}{{\mathcal K}}

\renewcommand{\Y}{{\mathcal Y}}
\renewcommand{\P}{{\mathcal P}}
\renewcommand{\U}{{\mathcal U}}
\renewcommand{\C}{{\mathscr C}}
\newcommand{\BH}{B^{\scriptscriptstyle \H}}
\newenvironment{oldthm}[1]{\par\noindent{\bf Theorem #1:} \em \noindent}{\par}
\newenvironment{oldlem}[1]{\par\noindent{\sc Lemma #1:} \em \noindent}{\par}
\newenvironment{oldcor}[1]{\par\noindent{\sc Corollary #1:} \em \noindent}{\par}
\newenvironment{oldpro}[1]{\par\noindent{\sc Proposition #1:} \em \noindent}{\par}
\newcommand{\othm}[1]{\begin{oldthm}{\ref{#1}}}
\newcommand{\eothm}{\end{oldthm} \smallskip}
\newcommand{\olem}[1]{\begin{oldlem}{\ref{#1}}}
\newcommand{\eolem}{\end{oldlem} \smallskip}
\newcommand{\ocor}[1]{\begin{oldcor}{\ref{#1}}}
\newcommand{\eocor}{\end{oldcor} \medskip}
\newcommand{\opro}[1]{\begin{oldpro}{\ref{#1}}}
\newcommand{\eopro}{\end{oldpro} \medskip}
\let\citeyear\cite
\newcommand{\acc}{\mathit{acc}}
\newcommand{\init}{\mathit{init}}

\begin{document}
\title{A Knowledge-Based Analysis of the Blockchain Protocol}
\fullv{
\author{
Joseph Y. Halpern%
\thanks{Supported in part by NSF grants IIS-0534064, IIS-0812045, 
IIS-0911036, and CCF-1214844, and by AFOSR grants 
FA9550-08-1-0438, FA9550-09-1-0266, and FA9550-12-1-0040,
and ARO grant W911NF-09-1-0281.}
\\
Cornell University\\
halpern@cs.cornell.edu 
\and Rafael Pass%
\thanks{
Supported in part by an Alfred P. Sloan Fellowship, a Microsoft New
Faculty Fellowship,  NSF Awards CNS-1217821 and CCF-1214844,
NSF CAREER Award CCF-0746990, AFOSR Award
FA9550-08-1-0197, AFOSR YIP Award FA9550-10-1-0093, BSF Grant 2006317,
and DARPA and AFRL under contract
FA8750-11-2-0211. The views and conclusions contained in this document
are those of the authors 
and should not be interpreted as the official policies, either
expressed or implied, of the Defense Advanced Research Projects Agency
or the US Government. }\\
Cornell University\\
rafael@cs.cornell.edu}

\maketitle
}

  \author{Joseph Y. Halpern
\institute{Cornell University\\
Ithaca, NY 14853, USA}
  \email{halpern@cs.cornell.edu}
\and
Rafael Pass
\institute{Cornell University\\
Ithaca, NY 14853, USA}
\email{rafael@cs.cornell.edu}
}
\shortv{
\def\titlerunning{A Knowledge-Based Analysis of the Blockchain Protocol}
\def\authorrunning{J. Y. Halpern \& R. Pass}
}

\maketitle
  
\begin{abstract}
    At the heart of the Bitcoin is a \emph{blockchain} protocol, a
    protocol  for achieving consensus on a public
ledger that records bitcoin transactions.
To the extent that a
blockchain protocol is used for applications such as contract signing and 
making certain transactions (such as house sales) public, we need to
understand what guarantees the protocol gives us in terms
of agents' knowledge.  Here, we provide a complete characterization
of agent's knowledge when running a blockchain protocol using a
variant of common knowledge that takes into account the fact that
agents can enter and leave the system, it is not known which agents are in
fact following the protocol (some agents may want to deviate if they
can gain by doing so), and the fact that the guarantees provided by
blockchain protocols are probabilistic.  We then consider some scenarios
involving contracts and show that this level of knowledge suffices for some
scenarios, but not others.
\end{abstract}


\section{Introduction}

At the heart of the Bitcoin \cite{nakamoto2008bitcoin} is
the \emph{blockchain} protocol, a protocol for achieving consensus on
a public 
ledger that records bitcoin transactions.
Indeed, much of the promise of Bitcoin involves using a blockchain
protocol
for applications that go far beyond a pure digital currency, such as 
an infrastructure for online payments, a way to record contracts and
asset exchanges, and a basis for dispute resolution.%
\footnote{Statistics showing Bitcoin's increasing usage can be found at
https://blockchain.info/charts;
even today, the size of the replicated ledger is over 50MB.
The article
``The great chain of being sure about things'' [\emph{The Economist},
  Oct. 31, 2015] provides a high-level discussion of potential
applications of blockchain protocols.}

At any given time when running a blockchain protocol, different agents
typically have different views about which transactions are in the
public ledger.  With current blockchain protocols, it is also possible that a
given transaction is included in agent $i$'s view of the ledger at
time $m$ and not included at a later time $m'$.  Nakamoto's protocol
\citeyear{GKL15,nakamoto2008bitcoin,PSS16} gives guarantees
in the spirit of the following, which we call \emph{$T$-consistency}
    (where $T$ is a 
non-negative integer):   Say that a ledger $X$ is
a \emph{$T$-prefix} of a ledger $Y$ if $X$ is any prefix of the
ledger that contains all but the last $T$ transactions in $Y$.
$T$-consistency says that if $i$ is honest 
(i.e., $i$ has followed the protocol since joining the system) 
and $X$ is a $T$-prefix of $i$'s blockchain at time $m$, then at
all times $m' \ge m$, all honest agents will have $X$ as a prefix of their
ledger.

Does $T$-consistency suffice to use a blockchain protocol for the
types of applications envisioned for it?
If not, what else do we need?
More generally, what guarantees do we get using a blockchain
protocol?
Of course, the answer to the latter question depends in part on the
application.  We focus here on contracts.  In the old days,  when
agents got together in one place to sign a contract, the fact that
the contract was in force was common knowledge: all agents knew that
all agents knew that all agents knew \ldots that the contract was in
force.  Today, with electronic signatures, we can get the same effect
if there is a global clock.  Suppose that the attorneys require that
signatures on the contract are received by 11:30 AM on the global clock
and, if they are, the contract will be in force at noon on the global
clock.  Then if signatures are indeed received by 11:30 AM and it is
common knowledge that messages from the attorney are all received
within at most 5 minutes, then at noon on the global clock all agents
know that at noon on the global clock all agents know \ldots that the
contract is in force.  That is, at noon, the agents have \emph{common
  knowledge} that the contract is in force. 

Can we get the equivalent common knowledge from $T$-consistency?
As we show here, $T$-consistency does not suffice.
Roughy speaking, the problem is the following: 
Suppose that at time $t$ agent 1 has 
the signed contract in a $T$-prefix of his ledger.  Thus, at
if 2 is honest. Unfortunately, the contract may not be in
a $T$-prefix of 2's ledger.  Moreover, if 2's ledger does not
grow, it may never be in a $T$-prefix of 2's ledger, so 2 will
never know that 1 knows about the contract.  

For agent 2 to know that the contract is on 1's ledger, agent 2's
ledger has to grow sufficiently long that the contract is
in
a $T$-prefix of 2's ledger.  Moreover, for 1 to have a bound on
the time by which he knows that 2 will have the contract in his ledger,
he must know that this growth will happen by a certain time.
That guarantee is provided by the property called
\emph{$\Delta$-weak growth} \cite{PSS16}, which says the following:
if $i$ is an honest agent and has a ledger of length $N$ at time
$t$, then all honest agents will have 
ledgers of length $N$ by time $t+\Delta$.  (Note that
$\Delta$-weak growth does not place any requirements on the content of the
ledger; it just talks about the length of the ledger.
$T$-consistency, on the other hand, does place requirements on the
content of ledgers.)  Here, we show that the combination of
$\Delta$-weak growth and $T$-consistency suffices not just for agent 1
to know that agent 2 will know (within time $\Delta$) that 1 will have the
contract in his ledger, the combination is
necessary and sufficient to achieve \emph{$\Delta$-$\Box$-common knowledge}
among the honest agents that the contract is in all of their ledgers.
Roughly speaking, $\Delta$-$\Box$-common knowledge 
\cite{HM90,HMW} of a formula $\phi$ holds among the honest agents if
each honest agent knows that within 
$\Delta$ all the honest agents will know from that point on that within
$\Delta$ all the honest agents will know from that point on \ldots $\phi$.


As shown in \cite{FHMV}, $\Delta$-common knowledge (everyone knows
within $\Delta$ that everyone knows within $\Delta$ \ldots) suffices
to ensure coordination among groups of agents within a time window of
$\Delta$.  In the context of contract signing, this means that once
the first person has signed the contract, every honest party in the
system will know within $\Delta$ time units that the contract has been
signed.  But because the set of honest agents is 
a
\emph{non-rigid} or \emph{indexical} set---its membership changes over
time---it does \emph{not} follow from $\Delta$-common knowledge that
new honest agents who enter the system will also know about the
contract being signed.  This does follow from $\Delta$-$\Box$--common
knowledge, which is why we want the stronger condition. 

Things are yet more complicated in our setting because the formula
$\phi$ of interest, being on the ledger, is also agent dependent;
a contract can be on 1's ledger without being on 2's ledger
Ignoring these subtleties, and just accepting for now that using 
a blockchain protocol gives $\Delta$-$\Box$-common 
knowledge, the question then arises whether this is what we need for
contracts.  
It is well known \cite{FHMV} that $\Delta$-common knowledge among a
fixed group $G$ of agents is
necessary and sufficient for coordination within a time window of $\Delta$.
That is, it is necessary and sufficient to guarantee that all agents
in $G$ perform a given action within a window of $\Delta$.  (Thus,
common knowledge is necessary and sufficient to guarantee simultaneous
coordination.)  $\Delta$-$\Box$-common knowledge turns out to be what
is needed to extend this result to an indexical set of agents.
$\Delta$-$\Box$-common knowledge (or just $\Delta$-common knowledge if
the set of agents is fixed), in turn, suffices for
time-stamped common knowledge if there is 
a global clock and a commonly-known upper bound on message-delivery
time.  Here, we show by example that while $\Delta$-$\Box$-common
knowledge suffices 
for some scenarios involving contracts, it does not suffice for all of
them.
In general, a few extra properties are needed (which are also
satisfied by some blockchain protocols).
These examples make it clear that we need a better understanding of
the properties needed for various applications.

Nakamoto's blockchain protocol \cite{nakamoto2008bitcoin} 
actually does not
quite provide $T$-consistency and $\Delta$-weak growth
\cite{GKL15,PSS16};
nor do subsequent blockchain protocols such as \cite{PS16a,PS16b}.
These protocols just
provide these properties with high probability.  Our results (and the
examples) for 
the deterministic case extend naturally to the probabilistic
case.



\section{Runs, systems, and knowledge: a review}

\paragraph{Runs and Systems:}
In order to reason about the knowledge of agents running a blockchain
protocol, we use the standard ``runs and systems'' framework
\cite{FHMV}, which we now briefly review.  The description below is
largely taken from \cite{FHMV}, which should be consulted for more
details and intuition.  However, some new subtleties arise because the
set of agents is not fixed.

A {\em multiagent system\/} consists of agents interacting over
time.  At each point in time, each agent is in some 
{\em local state\/}. Intuitively, an agent's
local state encapsulates all the information to which the agent has
access.  In the blockchain setting, we take agent $i$'s local
state at time $m$ to be $i$'s history: $i$'s initial state together
with the messages that agent $i$ has sent
and received (which together determine $i$'s ledger at time
$m$) and the outcome of 
random coin tosses, if the agent is randomizing.
This means that agents have
what is called \emph{perfect recall} \cite{FHMV}; roughly speaking,
they do not forget facts that they have learned.
It is also conceptually 
useful to have an ``environment'' whose state can be
thought of as encoding everything relevant to the description of the
system that may not be included in the agents' local states.  For
example, the environment state might include a list of the agents
currently in the system and which ones are honest; it could also
include the time on some clock that none of the agents have direct
access to.
The \emph{global state} of a system consists of the local state of
each agent currently in the system and the environment's state.
(We define global state more formally below.)

A global state describes the system at a given point in time.
But a system is not a static entity; it constantly changes.
Since we are mainly interested in how systems change over time,
we need to build time\index{time|(} into our model.
We define a {\em run\/}\index{run} to be a function from time to
global states.
Intuitively, a run is a complete description of how the system's
global state evolves over time.
For the purposes of this paper, we take time to range over the natural
numbers.   Thus, the initial global state of the system in a
possible execution~$r$ is $r(0)$,
the next global state is $r(1)$, and so on.

In general, there are many possible executions of a system:\ there could
be a number of possible initial states and many things that could happen from
each initial state.  In the case of a blockchain protocol, even
with a fixed initial state, there are different sets of agents who
could join or leave the system at various points and messages could
take different lengths of time to be delivered. 
Formally, a {\em system\/}\index{multi-agent system} is a nonempty set of runs.
Intuitively, these runs describe all the possible sequences of events
that could occur in the system.  Thus, we are essentially identifying a
system with its possible behaviors.\index{state!global|)}

We will be particularly interested in the set of runs generated when
the players are following a blockchain protocol $P$.
Formally, a protocol for agent $i$ is a function from $i$'s local states to
actions.  When we talk about a protocol like the blockchain protocol,
we implicitly have in mind a protocol for each agent.  We think of
the environment as running a protocol as well, which (among other things)
determines which agents act in each global state, and which messages will
be delivered.  We use the environment's protocol to model the
adversary's behavior. 
Since we are considering asynchronous systems, we allow
the adversary to delay the messages for an arbitrary amount of time.
We allow dishonest players, but we may want to limit the number of
dishonest players, or limit what they can do.  For simplicity, we
assume that all of these choices are under the control of the
environment.  Thus, for example, the environment decides which players
are going to be dishonest, when they become dishonest, and what they
do when they are dishonest.  
We typically think of the environment's protocol as being nondeterministic; for
example, the adversary can nondeterministically choose how long it
will take a message to be delivered or who is dishonest.
Following \cite{FHMV}, we take a \emph{context}
$\gamma$ to be a pair consisting of a protocol $P_e$ for the
environment and a set of possible initial global states that the
system can start in.  Given a context $\gamma$ and a protocol $P$ for
the agents, there is a system $\R_{P,\gamma}$ generated
by running protocol $P$ in context $\gamma$.  We refer the reader
to \cite{FHMV} for a more formal treatment.%
\footnote{In \cite{FHMV}, the context had other components.  We can
ignore these for our purposes here.}

In most previous work on multiagent systems, there are assumed to be a
fixed number $n$ of agents in the system, so for simplicity we take the
set of agents to be $\{1,\ldots, n\}$ and take a global state to have
the form $(s_e, s_1, \ldots, s_n)$, where $s_e$ denotes the
environment state and $s_i$ denotes agent $i$'s state.  However, as we observed
above, the set of agents in the systems we are interested in is an
indexical set, whose membership may change.  Thus, we can no longer
take global states to have this form.  We thus assume that there is a (possibly
infinite) set $\AG$ that contains the names of all agents that are
ever in the system.  Formally, an indexical set $\cS$ of agents in a system
$\R$ is a function from the points in $\R$ to subsets of $\AG$;
intuitively, $\cS(r,m)$ is the set of agents in $\AG$ who are in $\cS$
at the point $(r,m)$.  We will be particularly interested in two
indexical sets: $\A$, the set of agents currently in the system, 
and $\H$, the honest agents currently in the 
system, where $\H \subseteq \A$
(more precisely, $\H(r,m) \subseteq \A(r,m)$ for all points
$(r,m)$ in the system).  We then take $r(m)$, the global state at a
point $(r,m)$, to be the set $\{(s_i,i): i \in 
\A(r,m)\} \union \{(s_e,e)\}$; that is, $r(m)$ consists of the state
of each agent currently in the system tagged by the agent's name,
together with the environment state $s_e$ tagged with ``$e$'' for
``environment''.  (We assume that $e \notin \AG$, to avoid confusion.)

If $i \in \A(r,m)$ and $(s_i,i) \in r(m)$, then define $r_i(m) = s_i$.
We write $(r,m) \sim_i (r',m')$ if agent $i \in \A(r,m) \inter
\A(r',m')$ and $r_i(m) = r_i'(m')$; that is, $(r,m)
\sim_i (r',m')$ if $i$ is in the system at both $(r,m)$ and $(r',m')$
and $i$ has the same local state at both points.
In systems with a fixed set of agents, $\sim_i$ is an equivalence
relation on the points in $\R$; in our setting, $\sim_i$ is an
equivalence relation on  $\{(r,m): i \in \A(r,m)\}$.
Define $\K_i(r,m) = \{(r',m'): (r',m') \sim_i (r,m)\}$.  Note that if
$i \notin \A(r,m)$, then $\K_i(r,m) = \emptyset$.

\paragraph{Propositional and temporal reasoning in systems:}
Assume that we have a set $\Phi$ of primitive propositions whose truth
is determined by the global state.  In our intended applications, the
primitive propositions will be statements such as
``$X$ is a $T$-prefix of $i$'s ledger'' and
``agent $j$ is honest''.  An \emph{interpreted system} is a pair
$(\R,\pi)$ consisting of a sytem $\R$ and an \emph{interpretation}
$\pi$ that associates with each primitive $p$ and  global state $s$ a
truth value; that is, $\pi(p,s) \in \{\mathit{true},\mathit{false}\}$.
We can then define the truth of a Boolean combination of primitive
propositions at a point $(r,m)$ in an interpreted system $\I = (\R,\pi)$ in
the standard way:
$$\begin{array}{l}
(\I,r,m) \sat p \mbox{ for a primitive proposition $p$ iff }
  \pi(p,r(m)) = \mathit{true}\\
  (\I,r,m) \sat \phi \land \psi \mbox{ iff }
  (\I,r,m) \sat \phi \mbox{ and } (\I,r,m) \sat \psi\\
(\I,r,m) \sat \neg \phi  \mbox{ iff } (\I,r,m) \not\sat
  \phi
\end{array}$$

We can also reason about time using the standard temporal logic
operator $\Box$ and $\Circ^\Delta$:
$$\begin{array}{l}
(\I,r,m) \sat \Box \phi  \mbox{ iff }
(\I,r,m') \sat \phi \mbox{ for all $m' \ge m$}\\
(\I,r,m) \sat \Circ^\Delta \phi  \mbox{ iff }
(\I,r,m+\Delta) \sat \phi.\end{array}$$

As usual, we say that a formula $\phi$ is \emph{valid} in $\I =
(\R,\pi)$ if $(\I,r,m) \sat \phi$ for all points $(r,m) \in \R\times \IN$.

\section{Blockchain properties}

A blockchain protocol constructs a \emph{distributed
  ledger}.  What this means is that each agent running a blockchain
protocol has a current view of the ledger, where a ledger is just a
sequence $(t_1, t_2, \ldots, t_N)$ of transactions.  The details of
the transactions are not relevant to our discussion here; for our
purposes, we can just assume that there is a (commonly known) set $T$
of possible transactions, and each element $t_i$ in the ledger is in
$T$.

We think of the ledger constructed by a blockchain protocol as being
a ``public'' ledger.  Since each agent running a blockchain protocol has
its own view of the ledger, and the set of agents running the protocol
can change over time, we need to explain more carefully what
``public'' means in this context.  Given a ledger $L = (t_1, \ldots,
t_N)$,  the \emph{length} of the ledger $L$, denoted $|L|$, is $N$.  A
prefix of $L$ has 
the form $(t_1, \ldots, t_{N'})$, with $N' \le N$.
A \emph{T-prefix} of $L$ is ledger of the form $(t_1, \ldots, t_{M})$,
with $M \le N-T$ (so is the empty sequence if $N \le T$).

Some agents might deviate from a blockchain protocol, especially if
they think it is advantageous to do so.  Say that an agent is
\emph{honest} at time $m$ if $i$ is an agent in the system at time $m$
and it has followed the protocol from the
time that it joined the system up to time $m$.   
Consider the following three properties
of a  run $r$:
\begin{itemize}
  \item ($T$-consistency:)
For all times $m$ and $m' \ge m$, 
if $i$ is honest
at time $m$ in run $r$, $L'$ is a $T$-prefix of $L_i(r,m)$, $i$ ledger
at time $m$ in run $r$,
    and  $j$ is honest at time $m'$, then $L'$ is a
        prefix of $L_j(r,m')$. 
    \item ($\Delta$-weak growth:)
For all times $m$ and $m' \ge m + \Delta$, 
if $i$ is honest at time $m$ in $r$ and $j$ is honest at time 
$m'$, then $|L_j(r,m')| \ge |L_i(r,m)|$. 
    \item {$T$-$\Delta$--acceptability}:
For all times $m$ and $m' \ge m$, if $i$ is honest at time $m$ in $r$, $L'$
is a $T$-prefix of $L_i(r,m)$, and $j$ is honest at time $m'+\Delta$,
then $L'$ is a $T$-prefix of $L_j(r,m'+\Delta)$. 
\end{itemize}

The following is almost immediate:
\pro\label{pro:acceptable}
If a run satisfies $T$-consistency and
$\Delta$-weak-growth, then it is
$T$-$\Delta$--acceptable.
\epro

\prf Suppose that a run $r$ satisfies $T$-consistency and
$\Delta$-weak-growth.  If $i$ is honest at time $m$ in $r$,
$L'$ is a 
$T$-prefix of $L_i(r,m)$, $m' \ge m$, and $j$ is honest at time $m' \ge
m + \Delta$, then by $T$-consistency, $L'$ is a prefix of $L_j(r,m')$; by
$\Delta$-weak-growth, $|L_j(r,m')| \ge |L_i(r,m)|$, so $L'$ is a
$T$-prefix of $L_j(r,m')$. \eprf

A protocol \emph{$P$ is $T$-consistent (resp., satisfies $\Delta$-weak
growth, is $T$-$\Delta$--acceptable)} in context $\gamma$ iff all runs
in $\R_{P,\gamma}$ satisfy $T$-consistency (resp., $\Delta$-weak
growth, $T$-$\Delta$--acceptability).  It follows immediately from 
Proposition~\ref{pro:acceptable} that if a protocol is $T$-consistent
and satisfies $\Delta$-weak growth in context $\gamma$, then it is
$T$-$\Delta$--acceptable in context $\gamma$.

There is no
known
blockchain protocol that is $T$-$\Delta$--acceptable; that
is, none guarantees the properties of $T$-consistency and
$\Delta$-weak growth.  However, 
as shown in \cite{GKL15,PSS16}, Nakamoto's blockchain protocol
guarantees that these properties hold with high probability (taken over
the runs of the protocol) in
appropriate contexts (roughly, under the assumption that a majority of the
players are honest, and that some systems parameters---specifically,
what is referred to as the ``mining hardness''---are appropriately set
as a function of the worst-case delay on the networks),
and hence is
$T$-$\Delta$--acceptable with high probability in those contexts.
We defer a discussion of protocols with probabilistic guarantees to
Section~\ref{sec:prob}. 

\commentout{
More precisely, we
are interested in the following property:
\dfn\label{dfn:acceptablepr}
A blockchain protocol $P$ is {$T$-$\Delta$-$\eps$--acceptable} 
in context $\gamma$
if it
satisfies the following property:
if $i$ is honest at time $m$, $m' \ge m+\Delta$, and $j$ is honest at
time $m'$, then $\Pr(\mbox{$L'$ is a $T$-prefix of $L_j(r,m'+\Delta)$} \mid
\mbox{$L'$ being a $T$-prefix of $L_i(r,m'+\Delta)$})  \ge 1-\eps$ ,
where $\Pr$ is the probability on runs of determined by $P$
and $\gamma$.
\edfn
}

\section{A temporal characterization of blockchain protocols}

We can already give a characterization of blockchain protocols,
without using knowledge.  The characterization involves statements
about honest agents.  
In the language, we have primitive propositions $i \in \H$
and $\Tprefix(X,L_i)$.   We take $\pi_P$ to be such that
$i \in \H$ is true at $(r,m)$ if $i \in \H(r,m)$ and
$\Tprefix(X,L_i)$ is true at $(r,m)$ if $X$ is
a $T$-prefix of $L_i(r,m)$.
Given a context $\gamma$, 
let $\I_{P,\gamma} = (\R_{P,\gamma},\pi_P)$.
Note that whether $P$ is $T$-$\Delta$--acceptable will, in general,
depend on the context and, more specifically, what the adversary is
allowed to do.

  \pro\label{thm:blockchainchar1}
If $r$ is a $T$-$\Delta$--acceptable run of a protocol $P$, then
$$(\I_{P,\gamma},r,m) \sat i \in \H \land \Tprefix(X,L_i) \rimp
\Circ^{\Delta} \Box 
(j \in \H \rimp \Tprefix(X,L_j)).$$
\epro

\prf This is almost immediate from the definition of
  $T$-$\Delta$--acceptability, so we omit the details here. \eprf  

\cor\label{blockchaincharcor}
$P$ is $T$-$\Delta$--acceptable
in context $\gamma$
iff  for all $i, j \in \AG$ and ledgers $X$, the formula
    $$i \in \H \land \Tprefix(X,L_i) \rimp  \Circ^{\Delta} \Box
(j \in \H \rimp \Tprefix(X,L_j))$$ is valid in $\I_{P,\gamma}$.
\ecor

We also want an analogous characterization of blockchain protocols that give
only probabilistic guarantees.  However, there are some subtleties involved in
dealing  with probability, so we defer this to Section~\ref{sec:prob}.

\section{$\Delta$-$\Box$--common knowledge and indexical sets}

While Corollary~\ref{blockchaincharcor} does give us a
characterization of blockchain protocols, it does not give a good
intuition regarding what assurances agents have when they run a
blockchain protocol.  To provide this, we need to add the agents' knowledge to
the picture.

The standard way to reason about 
the knowledge of agents is to add a modal operators
$K_i$ to the language, where $K_i \phi$ is read ``agent $i$ knows
$\phi$.''  As usual, we say that $K_i \phi$ holds at a point $(r,m)$
if $\phi$ holds at all points that $i$ can't distinguish from $(r,m)$:
$$(\I,r,m) \sat K_i \phi \mbox{ iff  $(\I,r',m') \sat \phi$ 
  for all $(r',m') \in \K_i(r,m)$. }$$

Given a fixed set $G$ of agents, we take common knowledge
among $G$ to hold if everyone in $G$ knows, everyone in $G$ knows that
everyone in $G$ knows, and so on.  We add operators $E_G$ and $C_G$ to
the language for ``everyone in $G$ knows'' and ``it is common
knowledge among the agents in $G$''.  Taking $E^{n+1}_G\phi$ to be an
abbreviation of $E_G E^n_G \phi$, we have
$$\begin{array}{l}
(\I,r,m) \sat E_G \phi \mbox{ iff } (\I,r,m) \sat K_i \phi \mbox{ for
    all $i \in G$}\\
(\I,r,m) \sat C_G \phi \mbox{ iff } (\I,r,m) \sat E^n_G \phi \mbox{ for
    all $n \ge 1$.}
  \end{array}$$

There are two ways in which $C_G$ is insufficient for our purposes.
For one thing, as is well known \cite{FHMV}, common knowledge is
closely related to simultaneous coordination; we cannot obtain common
knowledge in asynchronous systems, where there is no common clock;
we are interested in the asynchronous setting for blockchain
applications.  Thus, we must consider variants of common knowledge,
such as $\Delta$-common knowledge, that are attainable for some
appropriate $\Delta$ (at least, if
we can assure that clocks are synchronized reasonably closely, an
assumption that is quite plausible for our application domain).
Secondly, we will typically be interested in facts that are
($\Delta$-)common knowledge among the honest agents, an indexical
set.  So we need to define common knowledge relative to indexical sets
$\cS$.

In general, an agent in $\cS$ may not know that it is in $\cS$.
For example, an agent $i$ may not know if it is honest at time $m$;
perhaps some fault resulted in it not following the protocol at the
previous step.  It might seem that the obvious way to define
$\ES \phi$ is just as $\land_{i \in \cS} K_i \phi$, in analogy to the
way that $E_G \phi$ is defined.  As shown in \cite{FHMV,MT}, this
whether it is in $\cS$ (i.e., if it is not the case that $i \in \cS
\rimp K_i (i \in \cS)$  is valid; note that the latter condition
implies that $i \notin \cS \rimp K_i (i \notin \cS)$ 
is also valid).  Instead, following \cite{MT}, we define $\BS_i \phi$
to be an abbreviation 
for $K_i (i \in \cS \rimp 
\phi)$; that is, $\BS_i \phi$ holds if $i$ knows that if it is in $\cS$,
then $\phi$ holds.  Thus,
$$\begin{array}{l}
(\I,r,m)  \sat  \BS_i\phi
  \mbox{~iff~}
(\I,r',m') \sat \phi
           \mbox{~for all~} (r',m') \in \K_i(s)
           \mbox{~such that ~} i \in \cS(r',m'). 
\end{array}$$
We can now define $\ES\phi$
as $\land_{i\in\cS}\BS_i\phi$,
and $\CS \phi$ as $\land_{n \ge 1} \ES^n\phi$;
more precisely, $(\I,r,m) \sat \ES \phi$ iff $(\I,r,m) \sat \BS_i\phi$ for all
$i \in \cS(r,m)$ and 
$(\I,r,m) \sat \CS \phi$ iff $(\I,r,m) \sat \ES^n\phi$ for all
$n \ge 1$.
It is easy to check that $K_i\phi \rimp \BS_i \phi$ is valid and if
$i$ knows whether $i$ is in $\cS$, 
then $\ES \phi$ is equivalent to $\land_{i \in \cS} K_i \phi$.

We will be interested in some variants of common knowledge.  Since
they are all defined the same way, we give the general approach once
and for all.  Let $X$ be a sequence of modal operators.  Then we
define $C^X_\cS \phi$ to hold if $(X\ES)^n \phi$ holds for all $n \ge
1$, where $(X\ES)^1 \phi$ is just $X\ES \phi$ and $(X\ES)^{n+1}\phi$ is
$X\ES(X\ES)^n\phi$.  Clearly, $\CS$ is is $C^X_\cS$ where $X$ is the
empty sequence.  For $\Delta \ge 0$, $\Delta$-common knowledge is
$C^X_{\cS}$ where $X$ is $\Circ^\Delta$.%
\footnote{There are two differences between the presentation of
  $\Delta$-common knowledge here and that in \cite{FHMV}.  The first
  is that we define variants of common knowledge in terms of infinite
  conjunctions rather than in terms of fixed points.
  Secondly, in \cite{FHMV}, $\Delta$-common knowledge of $\phi$  is
  taken to hold at the point $(r,m)$ if there is an interval of size
  $\Delta$ such that for each agent $i$, $K_i \phi$ holds at some
  point in the interval.  The definition given here is the one given
in \cite{HM90}.  In the presence of perfect recall (which, as we have
  observed, holds in the systems that we consider), the two
  definitions can be shown to be equivalent.}

One reason for the interest in ($\Delta$-)common knowledge is because
of its tight connection to coordination.  As mentioned in the
introduction, $\Delta$-common knowledge is a necessary and 
sufficient condition for agents to coordinate within a
window of $\Delta$.   For the reasons discussed in the introduction,
we are interested in $\Delta$-$\Box$--common knowledge, that is,
$C^X_\cS$ where $X$ is $\Circ^\Delta \Box$. 

We want to prove that a formula of the form
$i \in \cS \land \psi \rimp C_{\cS}^X \psi$ is valid.
The standard way to prove that $\psi \rimp C_G \phi$ (for a fixed
group $G$) is valid is to show that $\psi \rimp C_G (\phi \land \psi)$.  
This is called the \emph{induction rule}.  As observed in
\cite[Exercise 6.13(d)]{FHMV}, the induction rule can also be used for
indexical common knowledge.  We want to apply it to indexical variants
of common knowledge. 
Say that $Y$ is a \emph{simple}
sequence of modal operators if there is a relation $\Y$ on points
such that $(\I,r,m) \sat Y\phi$ iff $(\I,r',m') \sat \phi$ for all
points $(r',m') \in \Y(r,m) = \{(r',m'): ((r,m),(r',m')) \in \Y\}$. 
Note that $\Circ^\Delta$ and $\Box$ are simple, and simple operators
are closed under composition, so $\Circ^{\Delta}\Box$ is simple.

\lem\label{lem:induction} If $Y$ is simple and 
$i \in \H \land \psi \rimp YE{_\H}(\phi \land
\psi)$ is valid for all $i \in \H$, then so is $i \in \H \land \psi
\rimp C^Y_\H(\phi)$. 
\elem

\shortv{We defer the proof of this and all later results to the full paper.}

The formula  $i \in \H \land \Tprefix(X,L_i) \rimp \Circ^{\Delta} \Box
(j \in \H \rimp \Tprefix(X,L_j))$ from
Proposition~\ref{thm:blockchainchar1} is actually not far from having the
form $i \in \cS \land \psi \rimp Y\ES(\phi \land \psi)$ needed to apply
Lemma~\ref{lem:induction}.  The antecedent of the
formula has the right form, and $Y$ is clearly $\Circ^{\Delta} \Box$,
which, as we have observed, is simple. It is also not hard to show that
$(j \in \H \rimp \Tprefix(X,L_j))$ implies $\BH_j(\Tprefix(X,L_j))$.
The only thing that prevents us from applying
Lemma~\ref{lem:induction} is that the arguments of the $B_j$ operators
are different formulas.  But they all say roughly the same thing:
$X$ is a $T$-prefix of ``my'' ledger.   We now modify the logic so
that the formulas say exactly this.

Specifically, we add the primitive propositions $\Tprefix(X,L)$ and $I
\in \H$ to the language.  The intended interpretation of the first formula is
just what we said above: ``$X$ is a $T$-prefix of my ledger''; the
intended interpretation of the second formula is ``I am honest''.
These are \emph{agent-relative} formulas; following \cite{Grove95,GroveH2},
we give such formulas semantics by having an agent on the left-hand side of
$\sat$ as well as $(\I,r,m)$.  
We have to redefine the semantics of all formulas in the more general
setting.  The semantics of conjunction and negation and of temporal
operators is unaffected, but the semantics of some of the primitive
propositions and of the knowledge operator is affected.  Specifically,
$$\begin{array}{ll}
(\I,r,m,i) \sat \Tprefix(X,L) \mbox{ iff $X$ is a $T$-prefix of
    $L_i(m)$}\\
  (\I,r,m,i) \sat I \in \H \mbox{ iff $i \in \H(r,m)$}\\
(\I,r,m,i) \sat \phi \land \psi \mbox{ iff }
  (\I,r,mi) \sat \phi \mbox{ and } (\I,r,m,i) \sat \psi\\
(\I,r,m,i) \sat \neg \phi  \mbox{ iff } (\I,r,m,i) \not\sat
  \phi\\
(\I,r,m,i) \sat \Box \phi  \mbox{ iff }
(\I,r,m',i) \sat \phi \mbox{ for all $m' \ge m$}\\
(\I,r,m,i) \sat \Circ^\Delta \phi  \mbox{ iff }
(\I,r,m+\Delta,i) \sat \phi\\
(\I,r,m,i) \sat K_j\phi  \mbox{ iff $(\I,r',m',j) \sat \phi$ for all
    $(r',m') \in \K_j(r,m)$}
\end{array}$$

Note that in the semantics for $K_j \phi$, we use $\K_j$ and
give the semantics relative to $j$.
Intuitively, this says that $j$ knows $\phi$ from $i$'s perspective if
$j$'s interpretation of $\phi$ is true in all worlds that $j$ considers
possible.  Although other choices are possible (see \cite{Grove95} for
discussion), this choice is the one that was adopted
in \cite{Grove95,GroveH2} and works well for our purposes.
The remaining clauses of the definition of $\sat$ are the same as
before; we omit the details here.  The agent $i$ just comes along for
the ride, so to speak, in the other clauses; it is only relevant in
giving semantics where who ``I'' is matters.   We continue to view
$\BS_i \phi$ as an abbreviation for $K_i (i \in \cS \rimp \phi)$ and 
$\ES\phi$ as an abbreviation for $\land_{i\in\cS}\BS_i\phi$.   The
definition $C^Y_\cS \phi$ is also unchanged.  Say that
$(\I,r,m) \sat \phi$ if $(\I,r,m,i) \sat \phi$ for all agents $i$.
As before, $\phi$ is valid in $\I$ if $(\I,r,m) \sat \phi$ for all
points $(r,m)$.

This language gives us just what we want.

  \thm\label{thm:blockchainchar} 
The following are equivalent:
  \begin{itemize}
\item[(a)]   $P$ is $T$-$\Delta$-acceptable in context $\gamma$;
\item[(b)] for all $i, j \in \AG$
 and ledgers $X$,
$$i \in \H \land \Tprefix(X,L_i) \rimp \Circ^{\Delta} \Box
(j \in \H \rimp \Tprefix(X,L_j))$$ is valid in $\I_{P,\gamma}$.
\item[(c)]
for all ledgers $X$,
  $I \in \H \land \Tprefix(X,L) \rimp  \Circ^{\Delta} \Box
E_\H (\Tprefix(X,L))$ is valid in $\I_{P,\gamma}$.
\item[(d)] for all ledgers $X$,
  $I \in \H \land \Tprefix(X,L) \rimp  
C^{\Circ^{\Delta} \Box}_{\H}(\Tprefix(X,L))$ is valid in $\I_{P,\gamma}$.
      \end{itemize}
\ethm

An immediate consequence of Theorem~\ref{thm:blockchainchar} is that
$T$-consistency does not suffice to get $\Delta$-$\Box$ common
knowledge; we really do need $\Delta$-weak growth.  (It is also not
hard to construct an explicit example showing this.)

\section{Adding probability to the framework}\label{sec:prob}
\commentout{
Typically when running randomized protocols, there is an obvious
probability on the set of runs in a system.
A {\em probabilistic system\/} is a pair
$(\R,\Pr)$, where $\R$ is a system (a set of runs) and $\Pr$ is a
probability on $\R$.  For simplicity, assume that every run in $\R$ is
measurable, so that $\Pr(r)$ is well defined for $r \in \R$.
}

To give semantics to questions like 
`What is the probability according to agent $i$
that transaction $t$ is in agent $j$'s ledger?'' at a point $(r,m)$,
agent $i$ needs a probability 
defined on points in $\K_i(r,m)$, the points that $i$ considers
possible at $(r,m)$.
Agent $i$'s probability 
of a formula $\phi$ at the point $(r,m)$ is then just the probability
of the set of points in $\K_i(r,m)$ where $\phi$ is true.

To define a probability on the points in $\K_i(r,m)$, we use the
approach suggested by Halpern and Tuttle \citeyear{HT}.
Given a protocol $P$ run in a context $\gamma$,
ideally, we would have a probability on the runs in
$R_{P,\gamma}$.  However, it may not be reasonable to assume a single
probability on the runs in $R_{P,\gamma}$, since that would
require a probability on the adversary's nondeterministic choices. 
The first step (quite standard in distributed computing) 
is to  factor out these choices so that, intuitively, there is only one
nondeterministic choice, and that is made at the first step---the
adversary chooses a deterministic or probabilistic protocol.  We then
partition the set 
of runs into a set $\C$ of cells, and assume that we have, for each
cell $C \in \C$, a probability $\mu_C$ on the runs in cell $C$.
Intuitively, each cell $C$ consists of the set of runs where the
adversary is using a particular probabilistic (or deterministic)
protocol.  Let $\C(r)$ denote the unique cell containing the run $r$.
We take a probabilistic interpreted system to be a
tuple $\I = (\R, \pi, \C, \{\mu_C\}_{C \in \C})$.  Given a
probabilistic protocol $P$ and a context $\gamma$, we assume that
$\gamma$ determines $\C$ and $P$ and $\gamma$ together
determine $\{\mu_C\}_{C \in \C}$, so that the probabilistic interpreted
system $\I_{P,\gamma}$ is well defined.

We want an analogue of  Theorem~\ref{thm:blockchainchar} for
probabilistic systems.  We first define a probabilistic analogue of
acceptability.
\dfn\label{dfn:acceptablepr}
A blockchain protocol $P$ is {$T$-$\Delta$-$\eps$--acceptable} 
in context $\gamma$, 
if, for all cells $C$ in $\I_{P,\gamma}$, 
$\mu_C(\{r \in C: \mbox{$r$ is $T$-$\Delta$--acceptable}\}) \ge 1-\eps$
(i.e., no matter what protocol the adversary is using, with probability
at least $1-\eps$, the probability that a run is
$T$-$\Delta$--acceptable is at least $1-\eps$).
\edfn

In Definition~\ref{dfn:acceptablepr},
$\mu_C$ is a probability on runs: that is,
appropriate properties hold with high probability taken on runs.  But
the analogue of Theorem~\ref{thm:blockchainchar} that we are
interested in 
considers
agents' beliefs at a point.  It is consistent that a
protocol $P$ is $T$-$\Delta$-$\eps$--acceptable, yet an honest agent
$i$ gets some information at a point $(r,m)$ that tells $i$ that $P$
is somehow compromised and
$r$ is not $T$-$\Delta$-$\eps$--acceptable.
While 
it is unlikely that $i$ gets such information, it is not impossible.
We deal with this using an idea that goes back to Moses and
Shoham \citeyear{MosesShoham}.  Let $\acc$ be a predicate on runs;
that is, $\acc(r)$ is either true or false for each run $r$.
Intuitively, we think of 
$\acc(r)$ as holding exactly if $r$ is $T$-$\Delta$--acceptable, but
we do not need to require this.
We will restrict attention to runs that satisfy $\acc$ in all our
definitions.  We abuse notation and also view $\acc$ as a primitive
proposition in the language, and take $\pi$ to be such that
$(\I,r,m) \sat \acc$ iff $\acc(r)$ holds.
We say that an interpretation $\pi$ is \emph{acceptable with respect
to $\R_{P,\gamma}$} if $\pi$ interprets $i \in \H$ and $\Tprefix(X,L_i)$
as discussed earlier, and interprets $\acc$ so that it depends only on
the run; that is, $\pi(\acc,r(m)) = \pi(\acc,r(0))$, so that $\acc$ is
true either at all points of a run or none of them.  Acceptable
interpretations can differ only in how they interpret $\acc$.

Define $B_i^{\cS,\acc}\phi$ to be an abbreviation of
$K_i(i \in \cS \land \acc \rimp
\phi)$ and $E_{\cS}^\acc\phi$ be an abbreviation of
$\land_{i \in \cS} B_i^{\cS,\acc} \phi$.  If $Y$ is a simple operator, let
$C^{Y,\acc}_{\cS}\phi$ be the infinite conjunction
$\land_{n\ge 1} (YE_{\cS}^\acc)^n\phi$.

Finally, add the formula $\init(\Pr(\phi) \ge \alpha)$ to the
language,
where if $\I = (\R, \pi, \C, \{\mu_C\}_{C \in \C})$, then
$(\I,r,m) \sat \init(\Pr(\phi) \ge \alpha)$ if
$\mu_{C(r)}(\{r \in C(r): (\I,r,0) \sat \Box \phi\}) \ge \alpha$.  Intuitively,
$\init(\Pr(\phi) \ge \alpha)$ is true at $(r,m)$ if the prior
probability of $\phi$ being always true is at least $\alpha$, given the
adversary is using the protocol determined by $C(r)$.

Now essentially the same
arguments as those used to prove Lemma~\ref{lem:induction}
and Theorem~\ref{thm:blockchainchar} can be used to prove the following
analogues of these results.

\lem\label{lem:inductionpr} If
$i \in \H \land \psi \land \acc \rimp \Circ^{\Delta}\Box E^\acc_{\H}(\phi \land
\psi)$ is valid for all $i \in \H$, then so is $i \in \H \land \psi
\rimp C^{\Circ^{\Delta}\Box,\acc}_{\H}\phi$. 
\elem


\thm\label{thm:blockchaincharpr} The following are equivalent:
  \begin{itemize}
\item[(a)]   $P$ is $T$-$\Delta$-$\eps$-acceptable in context $\gamma$;
\item[(b)] there is an interpretation $\pi$ acceptable for
    $\R_{P,\gamma}$ such that for all $i, j \in \AG$ and ledgers $X$, 
$$\init(\Pr(\acc) \ge 1-\eps) \land [i \in \H \land \Tprefix(X,L_i) \land \acc \rimp (\Circ^{\Delta}
\Box 
(j \in \H \rimp \Tprefix(X,L_j)))]$$
is valid in $(\R_{P,\gamma},\pi)$. 
\item[(c)]
there is an interpretation $\pi$ acceptable for $\R_{P,\gamma}$ such that
for all ledgers $X$,
$$\init(\Pr(\acc) \ge 1-\eps) \land [I \in \H \land \Tprefix(X,L) \land \acc \rimp  
\Circ^{\Delta}  \Box 
E_{\H}^{\acc} (\Tprefix(X,L)))]$$
is valid in $(\R_{P,\gamma},\pi)$.
\item[(d)]
there is an interpretation $\pi$ acceptable for $\R_{P,\gamma}$ such that
for all ledgers $X$,
$$\init(\Pr(\acc) \ge
1-\eps) \land [I \in \H \land \Tprefix(X,L) \land \acc \rimp  
C^{\Circ^{\Delta} \Box,\acc}_{\H}(\Tprefix(X,L))]$$
is valid in $(\R_{P,\gamma},\pi)$.
\end{itemize}
\ethm



\section{Discussion}
Our results provide a characterization
of two natural properties of blockchain protocols---$T$-consistency
\cite{GKL15,nakamoto2008bitcoin,PSS16}, and
$\Delta$-weak growth \cite{PSS16}---in terms of $\Delta$-common
knowledge of a
$T$-prefix of the ledger. What does this tell us in terms of what we
can use a blockchain protocol for?

First, note that neither consistency or growth tell us anything about
how player can \emph{add} content to a ledger. 
In \cite{GKL15,PSS16} an additional property, referred to
as \emph{$\Delta'$-liveness}, is defined, which, roughly speaking, stipulates
that if an honest 
player wants to add some message to the ledger, it will appear
there within $\Delta'$ time. 
We can easily characterize this property by adding appropriate 
primitive propositions to the language.

A more interesting question relates to when $\Delta$-common knowledge
suffices for applications such as, for example, contract signing. 
Consider a simple game-theoretic model to illustrate
some of the subtleties. We have two players, and a
third entity, the judge. For simplicity, assume that the system is
synchronous.  (We can easily extend these ideas to asynchronous
systems, but there are a number of minor subtleties that are
orthogonal to the main points we want to present, so we stick to
synchronous systems here.)  In each round, 
each player observes the contents of her
ledger and either signs the contract or waits.

The utility of the players is defined as follows:
\begin{itemize}
  \item If event $E$ happens on some $T$-prefix of the judge's ledger
(where, formally, an event $E$ is just a set of prefixes of ledgers,
and $E$ happens on a $T$-prefix $L'$ if $L' \in E$) 
and both players sign the contract within
$\tilde{\Delta} \geq 2\Delta$ steps of 
$E$ happening (on the judge's ledger) for the first time,
then both 
players get some ``high'' utility. 
\item If one player signs and the other does not, the
signing player gets utility $-\infty$, and the non-signer gets
  utility $0$. If nobody ever signs the contract, both players get
  utility $0$.
\item Finally, a player who signs the contract without 
    event $E$ happening on a $T$-prefix of the judge's ledger
within
  $\tilde{\Delta}$ steps gets utility $-\infty$.
\end{itemize}
Intuitively, the game models a situation where, based on the content of
a ledger (and in particular, whether the event $E$ happens on a
$T$-prefix 
of the ledger), players both want to sign a contract, but only
if 1) the event actually happened, and 2) \emph{both} players actually sign
fast enough after the event happening.

If this game is played in the presence of a blockchain protocol that satisfies
$T$-consistency and $\Delta$-growth, then it is clearly a Nash
equilibrium for players to sign the contract whenever
$E$ happens on some $T$-prefix of their ledger: by $\Delta$-weak
growth, the 
ledger  of the other player will be at least as long within $\Delta$
time, so by $T$-consistency, $E$ will also hold in his
$T$-prefix; finally, by $\Delta$-weak growth and $T$-consistency, this
could have happened at most $\Delta$ time ago for the judge. Thus,
both players will sign within $\Delta$ time of each other and this
must happen within $2\Delta$-steps of 
when
$E$ first happens on some $T$-prefix of the  judge's
ledger.
The key point is that when $E$ first happens on the
judge's ledger, it is $\tilde{\Delta}$-common
knowledge that it has happened and that, within at most $\tilde{\Delta}$, it
will be on some $T$-prefix of the judge's ledger.  It is easy to
check that this property suffices to guarantee that signing when they
know that $E$ has happened will give both players high utility.  There
is a sense in which this condition is necessary.  Suppose that we
build a contract-signing protocol on top of another protocol that
handles knowledge dissemination (which, for us, is a blockchain
protocol). Roughly speaking, this means that the contract-signing 
protocol does not affect the agents' knowledge about $E$. 
Then, if we assume that if $E$ has not yet happened, then both players
assign positive probability to $E$ never happening, the
knowledge-dissemination protocol has to guarantee this level of
knowledge for the contract-signing protocol to be able to guarantee
the agents high utility when $E$ does happen.  (We make this 
precise in the full paper.)

Note that in this game, the ``actions'' (i.e., whether to sign) in the
game are external to 
the blockchain protocol and utilities are defined based on these external
actions. If we had used a blockchain protocol that also satisfies
$\Delta'$-liveness, we could have defined utility only as a
function of the judge's ledger: instead of playing the action $S$,
the players get to interact with the ledger and can add content to
it; ``signing a contract'' now means  adding a digitally signed version of the 
contract to the ledger.  The judge gives both players high
utility if versions of the contract digitally signed by each of them
appear on the judge's ledger
within some appropriate time $\tilde{\Delta}'$ after $E$ first happens on
some $T$-prefix of the judge's ledger.

An alternative way to model this game
would be to instead require the digital signatures to arrive on the judge's
ledger within some $T'$
\emph{blocks} after event $E$ first happens. If a blockchain
protocol satisfies an
additional property referred to as the \emph{chain-growth
upper bound} \cite{PSS16}, 
which stipulates that length of a ledger cannot grow too fast (so
that the signed contract will not be prevented from appearing on the
judge's ledger within $T'$ blocks by being ``crowded out'' by
other transactions), then
the same argument also applies to such situations.  (It is
straightforward to also characterize this chain-growth 
upper bound property in  a logic with appropriate primitive propositions.)

An appealing
feature of the final model is that whether the contract is deemed ``successfully
signed'' is now itself a property of the (judge's) ledger, and
thus, by our result, whenever the successful signing happens, it
becomes $\Delta$-common knowledge, independent of the signing
strategy; in particular, there is no longer a 
need for the judge! 
One other point worth making: although we have considered a system
with only two agents, we may further want to require that if other
(honest) agents enter the system, they will also agree that the
contract has been signed (and the original two agents get $-\infty$ if
this is not the case).  In that case, we need $\Delta$-$\Box$ common knowledge,
not just $\Delta$-common knowledge.

As this discussion shows, consistency and growth are themselves not
sufficient for applications of blockchain protocols to contracts. Once
we add appropriate additional properties (such as liveness and a
chain-growth upper bound),  
we can use our characterization for non-trivial applications within
contract signing. We leave open the question of better understanding
the properties needed for different types of contracts being executed
using a blockchain protocol.

\section*{Acknowledgements}
We thank Ron van der Meyden for useful comments.
Halpern was supported in part by NSF grant 
CCF-1214844, AFOSR grant
FA9550-12-1-0040, and ARO grants W911NF-14-1-0017 and W911NF-16-1-0397.
Pass was supported in part by a Microsoft Research Faculty Fellowship, 
NSF CAREER Award CCF-0746990,
NSF grant CCF-1214844, AFOSR Award
FA9550-12-1-0040, and BSF Grant 2006317.

\shortv{\bibliographystyle{eptcs}}
\bibliography{z,joe}

\end{document}